\documentclass[%
 reprint,
 amsmath,amssymb,
 aps,
]{revtex4-2}

\usepackage{graphicx}
\usepackage{dcolumn}
\usepackage{bm}
\usepackage[pagebackref=false,colorlinks=true,linkcolor=acsblue,citecolor=olive,urlcolor=acsblue]{hyperref}
\usepackage[mathlines]{lineno}
\usepackage{slashed}
\usepackage{subfloat}
\usepackage{pgfplots,subfigure}
\usepackage[dvipsnames]{xcolor}
\usepackage{geometry}
\definecolor{acsblue}{RGB}{17,76,139}
\begin{document}
\fontsize{8}{9}\selectfont
\preprint{APS/123-QED}

\title{Two-body Dirac equation in DSR: results for fermion-antifermion pairs}

\author{Nosratollah Jafari}
\email{nosrat.jafari@fai.kz}
\affiliation{Fesenkov Astrophysical Institute, 050020, Almaty, Kazakhstan}
\affiliation{Al-Farabi Kazakh National University, Al-Farabi av. 71, 050040 Almaty, Kazakhstan}
\affiliation{Center for Theoretical Physics, Khazar University, 41 Mehseti Street, Baku, AZ1096, Azerbaijan}

\author{Abdullah Guvendi}
\email{abdullah.guvendi@erzurum.edu.tr (Corr. Auth.)}
\affiliation{Department of Basic Sciences, Erzurum Technical University, 25050, Erzurum, Türkiye}

\date{\today}

\begin{abstract}
\vspace{0.15cm}
\setlength{\parindent}{0pt}

{\footnotesize This study investigates a modified two-body Dirac equation in \(2+1\)-dimensional spacetime, inspired by Amelino-Camelia's doubly special relativity (DSR). We begin by deriving a covariant two-body Dirac equation that, in the absence of DSR modifications, reduces to a Bessel-type wave equation. Incorporating corrections from the chosen DSR model modifies this wave equation, yielding solutions consistent with established results in the low-energy regime. We demonstrate that the effects of DSR modifications become particularly pronounced at large relative distances. For a coupled fermion-antifermion pair, we derive the modified binding energy solutions. By accounting for first-order Planck-scale corrections, we show that the fine-structure constant \(\alpha\) behaves as an energy-dependent running parameter, given by \(\alpha_{\text{eff}}(E)/\alpha \approx 1 - \frac{E}{4E_p}\), where \(E_p\) is the Planck energy. Binding energy levels are computed using a first-order approximation of the DSR modifications, and the results are applied to positronium-like systems. Our model reveals that DSR modifications induce shifts in the binding energy levels. To the best of our knowledge, DSR-modified two-body equations have not been previously studied. This model is the first of its kind, opening new avenues for further research in this area.}

\end{abstract}

\maketitle


\section{Introduction}

\vspace{0.15cm}
\setlength{\parindent}{0pt}

Doubly Special Relativity (DSR) theories are nonlinear modifications of special relativity with two invariant scales: the speed of light \( c \) and the Planck energy \( E_p = \sqrt{\hbar\, c^5/G} \approx 10^{19} \, \text{GeV} \). Two main examples of DSR theories are the Amelino-Camelia DSR \cite{Amelino-Camelia:2000stu} and the Magueijo-Smolin (MS) DSR \cite{Magueijo:2001cr}. Modifications to the Dirac equation can be obtained in these DSR theories. One of us (NJ) derived the modified Dirac equation in the Amelino-Camelia DSR \cite{Jafari:2024bph}. This work demonstrated that the modified Dirac equation reduces to Schrödinger equations for both particles and antiparticles, each characterized by distinct masses. Additionally, M. Coraddu and S. Mignemi obtained the Dirac equations in the MS DSR \cite{Coraddu:2009sb}. The modified Dirac equation has also been derived in other phenomenological quantum gravity models, such as loop quantum gravity \cite{Bojowald:2007pc, Li:2022szn, Melo:2024gxl}, models with a minimal length \cite{HoffDaSilva:2020uov, AntonacciOakes:2013qvs, Menculini:2013ida}, and models incorporating the generalized uncertainty principle (GUP) \cite{Das:2010zf, Faizal:2014mfa, Hamil:2019knw}. The modified Dirac equation can likewise be found in extensions of the Standard Model (SME) \cite{Lehnert:2004ri, Ferreira:2006kg, Casana:2012yj, Melo:2024rec}. SME models aim to identify potential quantum gravity effects by extending the Standard Model of particle physics with possible extra terms \cite{Colladay:1996iz, Kostelecky:2009zp}. In nearly all cases of the modified Dirac equation listed above, the degree of derivatives in the modified differential equations has increased. To our knowledge, only in our previous paper (NJ) \cite{Jafari:2024bph} and in the work of M. Coraddu and S. Mignemi \cite{Coraddu:2009sb} do the modified differential equations remain first-order, although in the first order of the Planck length approximation. In this case, the primary modifications to the Dirac equation involve changes to the masses of the particle and antiparticle. Therefore, we conjecture that solving both the Dirac equation and many-body Dirac equations should not be challenging.

\vspace{0.15cm}
\setlength{\parindent}{0pt}

On the other hand, non-relativistic quantum mechanics operates within a single time framework, utilizing free Hamiltonians for individual particles and an interaction term, where wave functions depend on each particle's spatial position. The Dirac equation's emergence spurred attempts to formulate a unified two-body Dirac equation, with foundational efforts by Breit \cite{Breit:1929zz}, who combined two free Dirac Hamiltonians and an interaction potential similar to the Darwin potential in electrodynamics. Yet, complications arise with long-range interactions or high particle velocities due to retardation effects. Later, Bethe and Salpeter \cite{Salpeter:1951sz} extended the work through quantum field theory, where relative time effects necessitated approximations of instantaneous interactions for particle systems. Decades after, Barut introduced a fully covariant many-body Dirac equation incorporating spin algebra and the most general electric and magnetic potentials \cite{Barut1985DerivationON}. His approach involves a spin algebra based on Kronecker products of Dirac matrices, resulting in a \(16 \times 16\) matrix equation within \((3+1)\)-dimensional spacetime. Angular and radial components are separated using group theoretical techniques \cite{Barut1985RadialEF}, although solving the resulting \(16\) radial equations remains challenging, even for systems as fundamental as the hydrogen atom. This difficulty arises from the coupled nature of the radial equations, producing pairs of second-order wave equations that require careful analysis \cite{Barut1985RadialEF}. However, recent developments indicate that the Barut's equation can be completely solved in lower-dimensional systems or systems with specific dynamical symmetries, whether in flat or curved spacetime \cite{Guvendi:2022uvz, Guvendi:2024nmm, Guvendi:2023aor, Guvendi:2020kei, Guvendi:2023rex, Dogan:2023lil}. These advancements have significantly enhanced the practical applications of the Barut's equation, leading to noteworthy breakthroughs \cite{Guvendi:2024diq}.

\vspace{0.15cm}
\setlength{\parindent}{0pt}

Modified many-body equations, within the context of various quantum gravity approaches, may hold significant potential for bridging the gap between quantum mechanics and gravity, with the ultimate aim of unifying general relativity and quantum field theory. This potential is particularly compelling in light of the absence of a comprehensive theory of quantum gravity. In this study, we present a modified two-body Dirac equation in a \(2+1\)-dimensional spacetime, drawing inspiration from a specific model of DSR \cite{Amelino-Camelia:2000stu,Jafari:2024bph}. We investigate both interacting and non-interacting particle scenarios, seeking to deepen our understanding of fundamental physics. The manuscript is structured as follows: In Section \ref{MDE-1}, we revisit the modified Dirac equation and examine its non-relativistic limits. Section \ref{TBDE} introduces the unmodified, fully covariant two-body Dirac equation in flat \(2+1\)-dimensional spacetime, where we focus on solutions that exclude mutual particle interactions. In Section \ref{MTBDE-1}, we present the modified form of the two-body Dirac equation and solve it, comparing the DSR corrections with the unmodified case. In Section \ref{coupled-pair}, we analyze a coupled fermion-antifermion pair, exploring the impact of DSR modifications on fermion-antifermion systems, such as positronium-like structures. Finally, in Section \ref{final}, we provide a detailed summary and discussion of our results.

\section{A brief overview of the modified Dirac equation} \label{MDE-1}

The modified Dirac equation in Amelino-Camelia's DSR, in the \( O(E^2 /  E^{2}_p) \) approximation for a single particle, can be expressed as follows \cite{Jafari:2024bph}:
\begin{equation}
\left[ i \gamma^0 \frac{1}{c} \frac{\partial}{\partial t} + i \gamma^i \frac{\partial}{\partial x^i} \left(1 + \frac{i}{2E_p} \frac{\partial}{\partial t} \right) - m \right] \tilde{\psi} = 0. \label{MDE}
\end{equation}
By defining
\begin{equation}
\tilde{\psi}(\vec{x}, t) = \begin{pmatrix} \tilde{\chi} \\ \tilde{\eta} \end{pmatrix},
\end{equation}
we obtain the following set of coupled equations \cite{Jafari:2024bph}:
\begin{align}
\left[E - m \right] \tilde{\chi} &= \left(1 + \frac{E}{2E_p} \right) \vec{\sigma} \cdot \vec{p} \, \tilde{\eta}, \\
\left[E + m \right] \tilde{\eta} &= \left(1 + \frac{E}{2E_p} \right) \vec{\sigma} \cdot \vec{p} \, \tilde{\chi}.
\end{align}
In the nonrelativistic limit \( (p^2 c^2 / m^2 c^4) \ll 1 \), these equations yield the Schrödinger equations \cite{Jafari:2024bph}:
\begin{equation}
i \frac{\partial \tilde{\chi}'}{\partial t} \approx -\frac{1}{2m_+} \nabla^2 \tilde{\chi}',
\end{equation}
and
\begin{equation}
i \frac{\partial \tilde{\eta}'}{\partial t} \approx \frac{1}{2m_-} \nabla^2 \tilde{\eta}',
\end{equation}
where we have introduced $\tilde{\chi'}=\exp{(+imc^2t)}\tilde{\chi}$, and $\tilde{\eta'}=\exp({-im c^2 t})\tilde{\eta}$. Also, \( m^+ \) and \( m^- \) are the modified masses for the particle and antiparticle, given by \cite{Jafari:2024bph}
\begin{equation}
m^{\pm} = \frac{m}{1 \pm \frac{mc^2}{E_p}}. \label{partAntipar}
\end{equation}
In traditional special relativity and relativistic quantum mechanics, the Schrödinger equation represents the non-relativistic limit of the Dirac equation. Particles and antiparticles have equal masses due to the dispersion relation \( E = \pm \sqrt{m^2 + p^2} \). However, in the considered DSR model, this mass equality is modified, introducing a difference of \( \frac{mc^2}{E_p} \) between the particle and antiparticle masses. The modified masses (\ref{partAntipar}) may influence the results of many-body systems, as illustrated throughout this paper.

\section{Two-body Dirac equation}\label{TBDE}

In this section, we present the fully covariant two-body Dirac equation in (2+1)-dimensional spacetime, formulated in Cartesian coordinates \((t, x, y)\). The spacetime background is characterized by a metric with a negative signature, given by \(ds^2 = c^2dt^2 - dx^2 - dy^2\). Within this framework, the covariant two-body Dirac equation for a fermion-antifermion (\(f\overline{f}\)) pair takes the form \cite{Guvendi:2024diq}:
\begin{equation}
\begin{split}
&\left\lbrace \mathcal{H}^{f} \otimes \gamma^{t^{\overline{f}}}+ \gamma^{t^{f}}\otimes \mathcal{H}^{\overline{f}} \right\rbrace \Psi(x_{\mu}^{f},x_{\mu}^{\overline{f}})=0,\\
&\mathcal{H}^{f(\overline{f})}= \slashed{\nabla}_{\mu}^{f(\overline{f})} +i\tilde{m}\mathcal{I}_{2},\\
&\slashed{\nabla}_{\mu}^{f (\overline{f})}=\gamma^{\mu^{f (\overline{f})}}\left(\partial_{\mu}^{f (\overline{f})}+i\frac{e_{f(\overline{f})}\mathcal{A}^{f(\overline{f})}_{\mu}}{\hbar c} \right).\label{eq0}
\end{split}
\end{equation}
In this context, \(\tilde{m}\) is defined as \(mc/\hbar\), where \(m\) represents the rest mass of individual fermions, \(e\) is the electric charge, \(\mathcal{A}_{\mu}\) denotes the 3-vector potential, and \(\hbar\) is the reduced Planck constant. Greek indices refer to coordinates in spacetime \((x^{\mu} = t, x, y)\), and \(\Psi(x^{\mu^{f}}, x^{\mu^{\overline{f}}})\) indicates the bi-local spinor dependent on the spacetime position vectors \((x^{\mu^{f}}, x^{\mu^{\overline{f}}})\) of the particles. Furthermore, \(\mathcal{I}_2\) represents the \(2 \times 2\) identity matrix. In equation (\ref{eq0}), the space-dependent Dirac matrices \(\gamma^{\mu}\) are derived using the relation \(\gamma^{\mu} = e^{\mu}_{(a)} \gamma^{(a)}\), where \(e^{\mu}_{(a)}\) denotes the inverse tetrad fields and \(\gamma^{(a)}\) are the space-independent Dirac matrices. These matrices, labeled \(\gamma^{(a)}\), are expressed in terms of the Pauli matrices \((\sigma_{x}, \sigma_{y}, \sigma_{z})\) \cite{Guvendi:2024diq}:
\begin{equation}
\gamma^{(0)}=\sigma_{z},\quad \gamma^{(1)}=i\,\sigma_{x},\quad \gamma^{(2)}=i\,\sigma_{y},\label{eq1}
\end{equation}
where \(i\) represents the imaginary unit \((i=\sqrt{-1})\). In the framework of a (2+1)-dimensional metric with a negative signature, where the Minkowski tensor \(\eta_{(a)(b)}\) is defined as \(\eta_{(a)(b)} = \text{diag}(+,-,-)\), the inverse tetrad fields are determined by the following expression:
\begin{equation*}
e^{\mu}_{(a)}=g^{\mu\tau}e_{\tau}^{(b)}\eta_{(a)(b)}.
\end{equation*}
Here, we denote the contravariant metric tensor by \( g^{\mu\tau} \), while \( e_{\tau}^{(b)} \) represents the tetrad fields, derived from the relation \( g_{\mu\tau} = e_{\mu}^{(a)} e_{\tau}^{(b)} \eta_{(a)(b)} \), where \( g_{\mu\tau} \) is the covariant metric tensor. This allows us to derive the corresponding generalized Dirac matrices \( (\gamma^{\mu}) \) as follows \cite{Guvendi:2024diq}:
\begin{equation}
\gamma^{t^{f(\overline{f})}}=\gamma^{(0)}/c,\quad \gamma^{x^{f(\overline{f})}}=\gamma^{(1)},\quad \gamma^{y^{f(\overline{f})}}=\gamma^{(2)}.\label{eq2}
\end{equation}
To simplify the analysis, we set \(\mathcal{A}_{\mu} = 0\) and present the corresponding two-body Dirac equation as \(\hat{M}\Psi = 0\), where \(\hat{M}\) is given by:
\begin{equation}
\begin{split}
&\gamma^{t^{f}}\otimes\gamma^{t^{\overline{f}}}\left[\partial_{t}^{f}+\partial_{t}^{\overline{f}} \right]+i\tilde{m}\left[\mathcal{I}_{2}\otimes \gamma^{t^{\overline{f}}}+\gamma^{t^{f}}\otimes \mathcal{I}_{2}\right]\\
&+\gamma^{x^{f}}\partial_{x}^{f}\otimes \gamma^{t^{\overline{f}}}+ \gamma^{t^{f}}\otimes \gamma^{x^{\overline{f}}}\partial_{x}^{\overline{f}}\\
&+\gamma^{y^{f}} \otimes\gamma^{t^{\overline{f}}}\partial_{y}^{f}+\gamma^{t^{f}}\otimes \gamma^{y^{\overline{f}}}\partial_{y}^{\overline{f}}\label{eq4}
\end{split}
\end{equation}
In accordance with the standard methodology for analyzing \(f\overline{f}\) systems, we define the coordinates for relative motion and center of mass motion as follows \cite{Guvendi:2024diq}:
\begin{equation*}
\begin{split}
&R_{x^{\mu}}=\frac{x^{\mu^{f}}}{2}+\frac{x^{\mu^{\overline{f}}}}{2},\quad r_{x^{\mu}}=x^{\mu^{f}}-x^{\mu^{\overline{f}}},\\
&x^{\mu^{f}}=\frac{1}{2}r_{x^{\mu}}+R_{x^{\mu}},\quad x^{\mu^{\overline{f}}}=-\frac{1}{2}r_{x^{\mu}}+R_{x^{\mu}},\\
&\partial_{x_{\mu}}^{f}=\partial_{r_{x^{\mu}}}+\frac{1}{2}\partial_{R_{x^{\mu}}},\quad \partial_{x_{\mu}}^{\overline{f}}=-\partial_{r_{x^{\mu}}}+\frac{1}{2}\partial_{R_{x^{\mu}}}, \label{eq5}
\end{split}
\end{equation*}
for a \(f\overline{f}\) pair. It is important to note that the combination \(\partial_{x^{\mu}}^{f} + \partial_{x^{\mu}}^{\overline{f}}\) simplifies to \(\partial_{R_{x^\mu}}\). This indicates that the system's time evolution is closely linked to proper time, represented by \(R_{t}\). We can now express the spacetime-dependent bi-spinor \(\Psi(t,\vec{r},\vec{R})\) in a factorized form:
\begin{equation*}
\Psi=e^{-i\,\frac{E}{\hbar}\, t}e^{i\vec{K} \cdot \vec{R}}\tilde{\Psi}(\vec{r})
\end{equation*}
where \(\tilde{\Psi}(\vec{r})=(\psi_{1}(\vec{r})\,\psi_{2}(\vec{r})\,\psi_{3}(\vec{r})\,\psi_{4}(\vec{r}))^{T}\). Here, \(E\) represents the relativistic energy, while \(\vec{r}\) and \(\vec{R}\) denote the spatial position vectors for relative motion and center of mass motion, respectively. The vector \(\vec{K}\) signifies the center of mass momentum, and \(^{T}\) indicates the transpose of the \(\vec{r}\)-dependent spinor. By assuming that the center of mass is stationary at the spatial origin, we can derive a set of equations that describe the relative motion of the pair in the center of mass frame, where the total momentum \((\vec{K})\) of the system is zero. Thus, we obtain
\begin{equation}
\begin{split}
&- \left( \gamma^t \otimes \gamma^t \right) i \varpi \tilde{\Psi}(\vec{r}) 
+ i \tilde{m} \left( I_2 \otimes \gamma^t + \gamma^t \otimes I_2 \right) \tilde{\Psi}(\vec{r})\\
&+ \left( \gamma^x \otimes \gamma^t - \gamma^t \otimes \gamma^x \right) \partial_x \tilde{\Psi}(\vec{r})\\
&+ \left( \gamma^y \otimes \gamma^t - \gamma^t \otimes \gamma^y \right) \partial_y \tilde{\Psi}(\vec{r}) = 0,\label{eq6}
\end{split}
\end{equation}
where \(\varpi=\frac{E}{\hbar\,c}\). By multiplying Equation (\ref{eq6}) on the left by \(i[\gamma^t \otimes \gamma^t]\)\footnote{It is important to note that \(c^2\left( \gamma^t \otimes \gamma^t \right)^2\) yields a \(4 \times 4\) dimensional identity matrix}, we can obtain the following \(4 \times 4\) dimensional matrix equation:
\begin{equation}
\begin{pmatrix}
\varpi - \tilde{\mu} & \hat{\partial}_- & -\hat{\partial}_- & 0 \\
-\hat{\partial}_+ & \varpi & 0 & -\hat{\partial}_- \\
\hat{\partial}_+ & 0 & \varpi & \hat{\partial}_- \\
0 & \hat{\partial}_+ & -\hat{\partial}_+ & \varpi + \tilde{\mu}
\end{pmatrix}\begin{pmatrix}
\psi_{1}(x,y) \\
\psi_{2}(x,y)\\
\psi_{3}(x,y)\\
\psi_{4}(x,y)
\end{pmatrix}= 0,\label{eq6a}
\end{equation}
where 
\[\tilde{\mu}= \frac{2mc}{\hbar}, \quad \hat{\partial}_\pm = \partial_x \pm i \partial_y.\]
\begin{figure}[h!]
\centering
\includegraphics[scale=0.40]{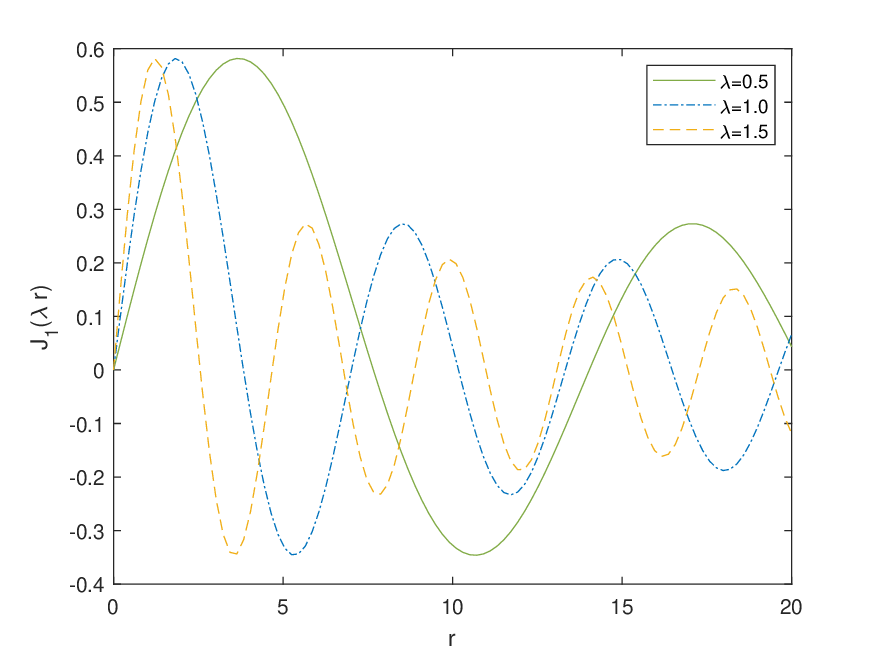}
\caption{This figure illustrates the oscillatory behavior of the Bessel function of the first kind, \( J_1(\lambda r) \), across a range of \( r \) values (from 0 to 20) for different arbitrary values of \(\lambda\) (0.5, 1, and 1.5). The curves demonstrate the characteristic oscillations and decay of the Bessel function, highlighting its dependence on the scaling factor \(\lambda\). Each curve is labeled with its corresponding \(\lambda\) value for clarity.}
\label{fig:1}
\end{figure}
We can now construct the possible spin states of an \(f \overline{f}\) pair. To achieve this, we utilize the following expressions: \(\hat{\partial} _{\mp} = e^{\mp i \phi} \left(\frac{d}{d r} \mp \frac{i}{r} \frac{d}{d \phi}\right)\). Here, \(\hat{\partial}_{+}\) and \(\hat{\partial}_{-}\) denote the spin raising and lowering operators, respectively \cite{Guvendi:2020kei}, and \(r\) represents the relative radial distance between the particles. This formulation leads to a set of coupled equations for the components of the transformed spinor \cite{Guvendi:2020kei},
\begin{equation*}
\tilde{\Psi}(r,\phi) \Rightarrow 
\begin{pmatrix}
\psi_1(r) e^{i(s-1)\phi} \\
\psi_2(r) e^{is\phi} \\
\psi_3(r) e^{is\phi} \\
\psi_4(r) e^{i(s+1)\phi}
\end{pmatrix},
\end{equation*}
as follows:
\begin{equation}
\begin{split}
&\varpi \varphi_1(r) - \tilde{\mu} \varphi_2(r) + 2 \frac{d \varphi_3(r)}{d r}= 0,\\
&\varpi \varphi_2(r) - \tilde{\mu}\varphi_1(r) + \frac{2s}{r} \varphi_3(r) = 0,\\
&\varpi \varphi_3(r) + \frac{2s}{r} \varphi_2(r) - 2 \left(\frac{1}{r} + \frac{d}{d r}\right) \varphi_1(r) = 0,\\
&\varpi \varphi_4(r) = 0,\label{eq7}
\end{split}
\end{equation}
where
\begin{equation*}
\begin{split}
&\varphi_1(r) = \psi_1(r) + \psi_4(r), \quad \varphi_2(r) = \psi_1(r) - \psi_4(r),\\
&\varphi_3(r) = \psi_2(r) - \psi_3(r), \quad \varphi_4(r) = \psi_2(r) + \psi_3(r).
\end{split}
\end{equation*}
Two of these equations simplify to algebraic form. In Equation (\ref{eq7}), \(s\) represents the total spin of the composite system formed by a \(f \overline{f}\) pair. It can be noted that \(\varphi_4(r)=0\) when \(\varpi\neq 0\), which leads to the conclusion that \(\varphi_3(r)=2\psi_2(r)\). Consequently, we can express the system of equations in the following manner:
\begin{equation}
\begin{split}
&\varpi \varphi_1(r) - \tilde{\mu} \varphi_2(r) + 4 \frac{d \psi_2(r)}{d r}= 0,\\
&\varpi \varphi_2(r) - \tilde{\mu}\varphi_1(r) = 0,\\
&\varpi \psi_2(r) -  \frac{\varphi_1(r)}{r} - \frac{d \varphi_1(r)}{d r}  = 0,
\end{split}
\end{equation}
if \(s=0\). This system of equations can be solved for \(\varphi_1(r)\), resulting in the following wave equation:
\begin{equation}
r^2 \frac{d^2 \varphi_1}{dr^2} + r \frac{d \varphi_1}{dr} + \left( \frac{\varpi^2 - \tilde{\mu}^2}{4} r^2 - 1  \right) \varphi_1 = 0.\label{WE}
\end{equation}
This equation is expressed in the standard Bessel form, characterized by the parameters \(\lambda = \sqrt{\frac{\varpi^2 - \tilde{\mu}^2}{4}}\) and \(\tilde{\nu} = 1\):
\begin{equation*}
r^2 \frac{d^2 \varphi_1(r)}{dr^2} + r \frac{d \varphi_1(r)}{dr} + \left( \lambda^2 r^2 - \tilde{\nu}^2 \right) \varphi_1(r) = 0.
\end{equation*}
The general solution to this equation is:
\begin{equation}
\varphi_1(r) = A\, J_1(\lambda r) + B\, Y_1(\lambda r),\label{MB}
\end{equation}
where \(A\) and \(B\) are arbitrary constants determined by the boundary conditions. The function \(J_1(\lambda r)\) oscillates for real values of \(r\) and has a series expansion that converges for all \(r\) (see also Fig. \ref{fig:1}).

\section{Modified two-body Dirac equation}\label{MTBDE-1}

\begin{figure}[h!]
\centering
\includegraphics[scale=0.40]{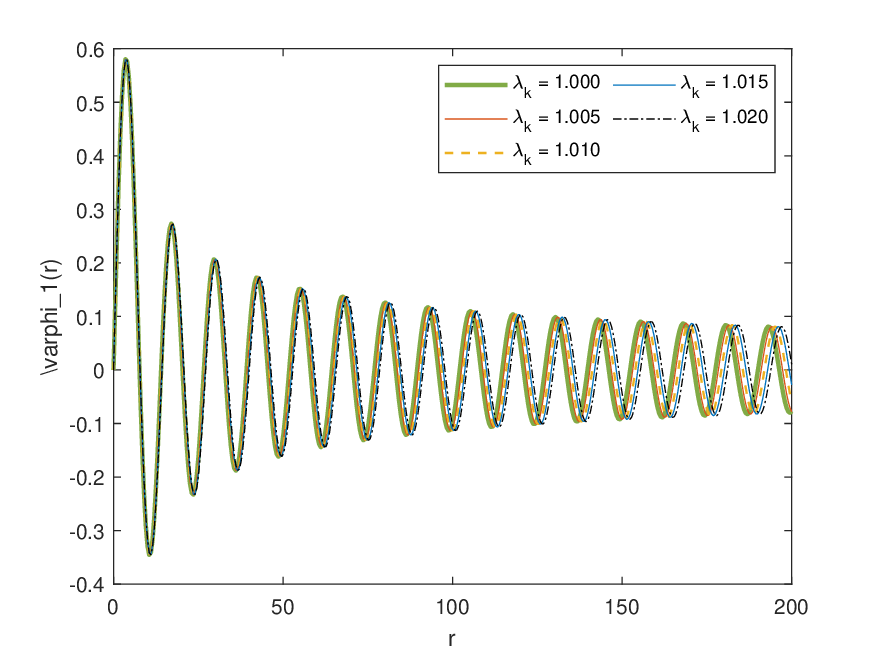}
\caption{\footnotesize This figure shows the behavior of the Bessel function-based solution \(\varphi_1(r)\) across the range \( r = [0, 200] \) for five different values of \(\lambda_k\): 1, 1.005, 1.010, 1.015 and 1.020. Each curve represents the oscillatory solution \(\varphi_1(r)\), with the amplitude adjusted by the factor \(\frac{\lambda}{\lambda_k}\), illustrating how slight changes in \(\lambda_k\) influence the oscillations in \(\varphi_1(r)\). The plot provides insight into the sensitivity of the solution to variations in \(\lambda_k\).}
\label{fig:2}
\end{figure}
In this section, we present the modified two-body Dirac equation. For two particles, based on Eqs. (\ref{MDE}) and (\ref{eq0}), the modified form can be expressed as follows:
\begin{widetext}
\begin{equation}
\left\lbrace \left[ \gamma^{t^{f}}  \partial^{f}_t + \gamma^{a^{f}} \partial^{f}_{a} \left(1 + \frac{i}{2E_p} \partial_t^{f} \right) + \tilde{m}\mathcal{I}_2 \right] \otimes \gamma^{t^{\overline{f}}}+ \gamma^{t^{f}}\otimes \left[ \gamma^{t^{\overline{f}}}  \partial^{\overline{f}}_t + \gamma^{a^{\overline{f}}} \partial^{\overline{f}}_{a} \left(1 + \frac{i}{2E_p} \partial_t^{\overline{f}} \right) + \tilde{m}\mathcal{I}_2 \right] \right\rbrace \Psi(x_{\mu}^{f},x_{\mu}^{\overline{f}})=0,\label{MTBDE}
\end{equation}
\end{widetext}
where \(a=x,y\). By applying the methodology outlined in Section \ref{TBDE}, we derive the following matrix equation:
\begin{equation}
\begin{pmatrix}
\varpi - \tilde{\mu} & \hat{\mathcal{D}}_- & -\hat{\mathcal{D}}_- & 0 \\
-\hat{\mathcal{D}}_+ & \varpi & 0 & -\hat{\mathcal{D}}_- \\
\hat{\mathcal{D}}_+ & 0 & \varpi & \hat{\mathcal{D}}_- \\
0 & \hat{\mathcal{D}}_+ & -\hat{\mathcal{D}}_+ & \varpi + \tilde{\mu}
\end{pmatrix}\begin{pmatrix}
\psi_{1}(x,y) \\
\psi_{2}(x,y)\\
\psi_{3}(x,y)\\
\psi_{4}(x,y) &
\end{pmatrix}= 0,\label{M-ME}
\end{equation}
where 
\begin{equation*}
\hat{\mathcal{D}}_{\pm}=\lambda_{k}\,\hat{\partial}_{\pm},\quad \lambda_{k}=\left[1+\frac{E}{4E_{p}}\right].
\end{equation*}
This matrix equation demonstrates how the DSR modifications influence the relative motion of an \( f\overline{f} \) pair. Consequently, by incorporating the effects of the DSR modifications, Eq. (\ref{WE}) is revised as follows:
\begin{equation}
\frac{d^2 \varphi_1}{dr^2} + \frac{1}{r} \frac{d \varphi_1}{dr} + \left( \frac{\varpi^2 - \tilde{\mu}^2}{4\lambda^{2}_{k}}  - \frac{1}{r^2} \right) \varphi_1 = 0.\label{M-WE}
\end{equation}
It is clear that the solution function is expressed as
\begin{equation}
\varphi_1(r) = A\, J_1\left(\frac{\lambda}{\lambda_{k}}\, r\right).
\end{equation}
This result demonstrates that the oscillation amplitude is influenced by \(\lambda_{k}\), emphasizing the role of DSR modifications in the behavior of the wave function, particularly at larger values of \(r\). This effect is depicted in Fig. \ref{fig:2}. Moreover, the model introduced in Ref. \cite{Guvendi:2024diq} shows that well-known systems can be studied, without loss of generality, within a position-dependent mass framework, \(\tilde{\mu} \rightarrow \tilde{\mu}(r)\). This modification effectively incorporates familiar systems, such as positronium and quarkonium.

\section{Effects of DSR modification on a coupled fermion-antifermion pair}\label{coupled-pair}

In this section, we seek to obtain the exact solution to the DSR-modified two-body Dirac equation within a position-dependent mass framework. To achieve this, we introduce a position-dependent mass modified by a Coulomb-type potential, defined as \(\tilde{\mu}(r) = \frac{2mc}{\hbar} - \frac{\alpha}{r}\) \cite{Guvendi:2024diq}. This enables us to rewrite the wave equation (\ref{M-WE}) as follows:
\begin{equation}
\frac{d^2 \varphi_1}{dr^2} + \frac{1}{r} \frac{d \varphi_1}{dr} + \left( -\tilde{a} + \frac{\tilde{b}}{r} - \frac{\tilde{c}}{r^2} \right) \varphi_1 = 0,\label{M-WE-S}
\end{equation}
where
\begin{equation*}
\tilde{a} = \frac{\tilde{\mu}^2 - \varpi^2}{4\lambda^2_k},\quad \tilde{b} = \frac{\tilde{\mu}\alpha}{2\lambda^2_k},\quad \tilde{c} = 1 + \frac{\alpha^2}{4\lambda^2_k},\quad \tilde{\mu} = \frac{2mc}{\hbar}.\label{parameters}
\end{equation*}
By introducing a new variable defined as \( z = 2\sqrt{\tilde{a}}\,r \), we can reformulate the wave equation. Utilizing the transformation \(\varphi_1(z) = \frac{\varphi(z)}{\sqrt{z}}\), the equation can be expressed as follows:
\begin{equation}
\begin{split}
&\frac{d^2 \varphi(z)}{dz^2} + \left(-\frac{1}{4} + \frac{\tilde{q}}{z} - \frac{\frac{1}{4} - \tilde{p}^2}{z^2}\right) \varphi(z) = 0,\\
&\tilde{q} = \frac{\tilde{b}}{2\sqrt{\tilde{a}}},\quad \tilde{p} = \sqrt{\tilde{c}}.\label{WEq-2}
\end{split}
\end{equation}
The solution to this wave equation near the regular singular point \( z = 0 \) can be expressed using the confluent hypergeometric function:
\begin{equation}
\varphi(z) = z^{\frac{1}{2} + \tilde{p}} e^{-\frac{z}{2}} \, {}_1F_1\left( \frac{1}{2} + \tilde{p} - \tilde{q}, 1 + 2 \tilde{\nu}; z \right).
\end{equation}
Thus, \(\varphi_1(z)\) can be easily determined using the expression \(\varphi_1(z) = \varphi(z)/\sqrt{z}\). For the confluent hypergeometric function \( {}_1F_1(q_1, q_2; z) \) to reduce to a polynomial in \( z \), it is both necessary and sufficient for the first parameter \( q_1 \) to be a non-positive integer. In the context of solving the wave equation, we define \( q_1 \) as \( \frac{1}{2} + \tilde{p} - \tilde{q} \) and impose the condition \( q_1 = -n \), where \( n \) is a non-negative integer (\( n \in \mathbb{Z}_{\geq 0} \)). This condition is crucial for determining the discrete energy levels of the system. Accordingly, we arrive at the following result:
\begin{equation}
\begin{split}
&E=\pm 2mc^2\sqrt{1-\frac{\alpha^2_{eff}(E)}{4\left[\tilde{n}+\sqrt{1+\frac{\alpha^2_{eff}(E)}{4}}\right]^2}},\\
&\tilde{n}=n+\frac{1}{2},\quad \alpha_{eff}(E)=\frac{\alpha}{1+\frac{E}{4E_p}}.\label{QG-ES}
\end{split}
\end{equation}
In principle, we can express this energy equation by assuming \(\alpha_{\text{eff}} \ll 1\) and employing the series expansion method, as is common in typical Coulomb problems:
\begin{equation}
\begin{split}
&E_n\approx\pm 2mc^2\left\{ 1-\frac{\alpha^2_{\text{eff}}(E)}{8(n+1)^2}+\frac{ \alpha^4_{\text{eff}}(E) (2n + 1) }{128 (n + 1)^4}-\mathcal{O}(\alpha_{\text{eff}}^6)\right\}.
\label{QG-ES-series}
\end{split}
\end{equation}
The first term represents the total rest mass energy (\(2mc^2\)) of fermion-antifermion pairs, while the second term corresponds to the well-known non-relativistic binding energy (\(E^{b}_{n}\)) for such a pair interacting via an attractive Coulomb potential, specifically in the limit where \(\frac{E}{E_{p}} \rightarrow 0\). By disregarding the DSR modifications, we derive the following energy spectrum for the binding energy of a non-relativistic singlet (\(s=0\)) positronium-like system:
\begin{equation}
E^{b}_{n}\approx -\,mc^2\,\frac{\alpha^2}{4(n+1)^2}.\label{QG-ES-usual-1}
\end{equation}
Assuming \( m \) represents the standard electron mass (\( m_e \)) and \( \alpha \approx 1/137 \), we calculate the ground state (\( n=0 \)) binding energy of a singlet positronium as follows \(
E^{b}_{0} \approx - m_{e}c^2 \alpha^2/4 \sim -6.801 \, \text{eV}\). \\

The result in equation \eqref{QG-ES-series} clearly indicates that the coupling parameter (\(\alpha\)) cannot be treated as a constant in an effective manner, as it varies with the energy scale of the system due to the effects of the DSR modification. Specifically, we find that \(\alpha\) is a running coupling parameter that depends on \(E\), expressed as \(\alpha_{eff}(E)/\alpha \approx 1 - \frac{E}{4E_p}\). It is important to note that the modified equations Eq.(\ref{MDE}) and Eq. (\ref{MTBDE}) incorporate first-order corrections due to DSR modification effects, allowing us to neglect higher-order terms \((E/E_p)^{d \geq 2} \sim 0\). In this framework, we can rewrite Eq. (\ref{QG-ES}) by approximating \(\alpha_{eff}(E)\) as \(\alpha \left(1 - \frac{E}{4E_p}\right)\), and using the approximation \(\left(1 - \frac{E}{4E_p}\right)^2 \approx 1 - \frac{E}{2E_p}\). Accordingly, the final simplified expression for the altered binding energy spectra is given by:
\begin{equation}
\begin{split}
&E^{b^{DSR}}_{n}= E^{b}_{n} \left( 1 - \frac{\mathcal{E}_{0}}{E_{p}} \right),\\
&\mathcal{E}_{0} = m_{e}c^2,\quad \frac{\mathcal{E}_{0}}{E_{p}} \approx 4.19 \times 10^{-23}.
\end{split}
\end{equation}
This result suggests that DSR modifications can, in principle, be "observed" even in the non-relativistic binding energy term (\(\propto \alpha^2\)). Furthermore, the magnitude of the non-relativistic binding energy decreases by a factor of \(\frac{\mathcal{E}_{0}}{E_{p}}\) for each quantum state (\(n\)). This conclusion is reinforced by the fact that \(\alpha_{\text{eff}} < \alpha\). The DSR modifications affect the binding energies across all quantum states, scaling the original energy levels by the same factor for each state. Here, it is evident that the binding energy shifts arising from the considered DSR model are \(|\Delta E^{b}_{n}|= E^{b}_{n}-E^{b^{DSR}}_{n}\). Therefore, we obtain the following expression for the shifts in the non-relativistic binding energy induced by DSR:
\begin{equation}
|\Delta E^{b}_{n}|\approx E^{b}_{n}\,\frac{\mathcal{E}_{0}}{E_{p}}= E^{b}_{n} \times 4.19 \times 10^{-23}\label{E-dif}
\end{equation}

\section{Summary and discussions}\label{final}

This research presents a comprehensive study of the two-body Dirac equation modified by DSR, inspired by Amelino-Camelia's DSR framework. Building upon this foundation, we introduce a novel two-body Dirac equation modified by the DSR model, set in a \(2+1\)-dimensional flat spacetime. This formulation enables a detailed analysis of both interacting and non-interacting particles. Our investigation begins with the formulation of a fully covariant two-body Dirac equation in \( (2+1) \)-dimensional spacetime characterized by a negative signature metric. Through our analysis, we derive a wave equation that adheres to the standard Bessel form. Next, we present the modified two-body Dirac equation, which incorporates the effects of DSR modifications on the \(f\overline{f}\) system. The solution demonstrates that the oscillation amplitude of the wave function, represented by the Bessel function \(J_1\), is influenced by the parameter \(\lambda_k\), which accounts for DSR effects. This finding indicates that the behavior of the wave function, particularly at larger radial distances, experiences "significant" alterations due to DSR corrections, as illustrated in Fig. \ref{fig:2}.

\vspace{0.10cm}
\setlength{\parindent}{0pt}

Finally, we investigate a coupled fermion-antifermion pair and derive the regular solution to the resulting wave equation in terms of the confluent hypergeometric function. In this context, we observe that the magnitude of the non-relativistic binding energy decreases by a factor of \(\frac{\mathcal{E}_{0}}{E_{p}}\) for each quantum state (\(n\)), reinforcing the conclusion that \(\alpha_{eff} < \alpha\). The effect of the considered DSR model corrections on binding energies across all quantum states reveals a consistent scaling of the original energy levels by the same factor for each state. Moreover, the magnitude of these DSR model corrections is very small, as demonstrated by our findings, approximately \(|\Delta E^{b}_{n}| \approx E^{b}_{n} \times \frac{\mathcal{E}_{0}}{E_{p}} = E^{b}_{n} \times 4.19 \times 10^{-23}\). Our results highlight the influence of the considered DSR model on many-body systems described by the two-body Dirac equation within a \( (2+1) \)-dimensional framework, showing that the considered DSR model induces small shifts in the binding energy.\\

\vspace{0.10cm}
\setlength{\parindent}{0pt}

Here, it is crucial to note that we consider only a specific model of DSR \cite{Amelino-Camelia:2000stu,Jafari:2024bph} in this research and calculate how these modifications affect the energy levels of a system of particle-antiparticle interactions via a Coulomb potential by obtaining the leading-order Planck-scale corrections. Within the DSR model we are considering, due to the lack of a complete framework capable of describing the dynamics of quantum particles in a context with deformed symmetries, our results should be regarded, at most, as a preliminary indication formulated in a simplified setting, where we have made several assumptions that may not necessarily remain valid in a fully developed framework.

\section{ Acknowledgment}
NJ has been funded by the Science Committee of the Ministry of Science and Higher Education of the Republic of Kazakhstan Program No. BR21881880.\\
The authors sincerely thank the reviewer for their thoughtful and constructive suggestions.




\nocite{*}

\bibliography{Refrences}

@article{Amelino-Camelia:2000stu,
    author = "Amelino-Camelia, Giovanni",
    title = "{Relativity in space-times with short distance structure governed by an observer independent (Planckian) length scale}",
    eprint = "gr-qc/0012051",
    archivePrefix = "arXiv",
    doi = "10.1142/S0218271802001330",
    journal = "Int. J. Mod. Phys. D",
    volume = "11",
    pages = "35--60",
    year = "2002"
}

@article{Jafari:2024bph,
    author = "Jafari, Nosratollah and Shukirgaliyev, Bekdaulet",
    title = "{Nonrelativistic limits of the Klein-Gordon and Dirac equations in the Amelino-Camelia DSR}",
    doi = "10.1016/j.physletb.2024.138693",
    journal = "Phys. Lett. B",
    volume = "853",
    pages = "138693",
    year = "2024"
}

@article{Coraddu:2009sb,
    author = "Coraddu, M. and Mignemi, S.",
    title = "{The Nonrelativistic limit of the Magueijo-Smolin model of deformed special relativity}",
    eprint = "0911.4241",
    archivePrefix = "arXiv",
    primaryClass = "hep-th",
    doi = "10.1209/0295-5075/91/51002",
    journal = "EPL",
    volume = "91",
    number = "5",
    pages = "51002",
    year = "2010"
}

@article{Bojowald:2007pc,
    author = "Bojowald, Martin and Das, Rupam and Scherrer, Robert J.",
    title = "{Dirac Fields in Loop Quantum Gravity and Big Bang Nucleosynthesis}",
    eprint = "0710.5734",
    archivePrefix = "arXiv",
    primaryClass = "astro-ph",
    reportNumber = "IGC-07-10-5",
    doi = "10.1103/PhysRevD.77.084003",
    journal = "Phys. Rev. D",
    volume = "77",
    pages = "084003",
    year = "2008"
}

@article{Li:2022szn,
    author = "Li, Hao and Ma, Bo-Qiang",
    title = "{Speed variations of cosmic photons and neutrinos from loop quantum gravity}",
    eprint = "2212.04220",
    archivePrefix = "arXiv",
    primaryClass = "hep-ph",
    doi = "10.1016/j.physletb.2022.137613",
    journal = "Phys. Lett. B",
    volume = "836",
    pages = "137613",
    year = "2023"
}

@article{Melo:2024gxl,
    author = {Melo, Joao Paulo S. and Neves, Mario J. and Paix\~ao, Jefferson M. A. and Helay\"el-Neto, Jos\'e A.},
    title = "{Loop quantum gravity effects on electromagnetic properties of charged leptons}",
    eprint = "2403.17197",
    archivePrefix = "arXiv",
    primaryClass = "hep-th",
    doi = "10.1140/epjc/s10052-024-13257-9",
    journal = "Eur. Phys. J. C",
    volume = "84",
    number = "9",
    pages = "938",
    year = "2024"
}

@article{HoffDaSilva:2020uov,
    author = "Hoff Da Silva, J. M. and Beghetto, D. and Cavalcanti, R. T. and Da Rocha, R.",
    title = "{Exotic fermionic fields and minimal length}",
    eprint = "2006.03490",
    archivePrefix = "arXiv",
    primaryClass = "hep-th",
    doi = "10.1140/epjc/s10052-020-8313-z",
    journal = "Eur. Phys. J. C",
    volume = "80",
    number = "8",
    pages = "727",
    year = "2020"
}

@article{AntonacciOakes:2013qvs,
    author = "Antonacci Oakes, T. L. and Francisco, R. O. and Fabris, J. C. and Nogueira, J. A.",
    title = "{Ground State of the Hydrogen Atom via Dirac Equation in a Minimal Length Scenario}",
    eprint = "1308.3395",
    archivePrefix = "arXiv",
    primaryClass = "hep-th",
    doi = "10.1140/epjc/s10052-013-2495-6",
    journal = "Eur. Phys. J. C",
    volume = "73",
    pages = "2495",
    year = "2013"
}

@article{Menculini:2013ida,
    author = "Menculini, L. and Panella, O. and Roy, P.",
    title = "{Exact solutions of the (2+1) Dimensional Dirac equation in a constant magnetic field in the presence of a minimal length}",
    eprint = "1302.4614",
    archivePrefix = "arXiv",
    primaryClass = "hep-th",
    doi = "10.1103/PhysRevD.87.065017",
    journal = "Phys. Rev. D",
    volume = "87",
    number = "6",
    pages = "065017",
    year = "2013"
}

@article{Das:2010zf,
    author = "Das, Saurya and Vagenas, Elias C. and Ali, Ahmed Farag",
    title = "{Discreteness of Space from GUP II: Relativistic Wave Equations}",
    eprint = "1005.3368",
    archivePrefix = "arXiv",
    primaryClass = "hep-th",
    doi = "10.1016/j.physletb.2010.05.052",
    journal = "Phys. Lett. B",
    volume = "690",
    pages = "407--412",
    year = "2010",
    note = "[Erratum: Phys.Lett.B 692, 342--342 (2010)]"
}

@article{Hamil:2019knw,
    author = "Hamil, B. and Merad, M.",
    title = "{Dirac Equation in the Presence of Minimal Uncertainty in Momentum}",
    doi = "10.1007/s00601-019-1505-0",
    journal = "Few Body Syst.",
    volume = "60",
    number = "2",
    pages = "36",
    year = "2019"
}

@article{Faizal:2014mfa,
    author = "Faizal, Mir and Kruglov, Sergey I.",
    title = "{Deformation of the Dirac Equation}",
    eprint = "1406.2653",
    archivePrefix = "arXiv",
    primaryClass = "physics.gen-ph",
    doi = "10.1142/S0218271816500139",
    journal = "Int. J. Mod. Phys. D",
    volume = "25",
    number = "01",
    pages = "1650013",
    year = "2015"
}

@article{Lehnert:2004ri,
    author = "Lehnert, Ralf",
    title = "{Dirac theory within the standard model extension}",
    eprint = "hep-ph/0401084",
    archivePrefix = "arXiv",
    doi = "10.1063/1.1769105",
    journal = "J. Math. Phys.",
    volume = "45",
    pages = "3399--3412",
    year = "2004"
}

@article{Ferreira:2006kg,
    author = "Ferreira, Jr., Manoel M. and Moucherek, Fernando M. O.",
    title = "{Influence of Lorentz- and CPT-violating terms on the Dirac equation}",
    eprint = "hep-th/0601018",
    archivePrefix = "arXiv",
    doi = "10.1142/S0217751X06033842",
    journal = "Int. J. Mod. Phys. A",
    volume = "21",
    pages = "6211--6227",
    year = "2006"
}

@article{Melo:2024rec,
    author = {Melo, Jo\~ao Paulo S. and Helay\"el-Neto, Jos\'e A.},
    title = "{Re-assessing special aspects of Dirac fermions in presence of Lorentz-symmetry violation}",
    eprint = "2404.03692",
    archivePrefix = "arXiv",
    primaryClass = "hep-ph",
    doi = "10.1016/j.aop.2024.169790",
    journal = "Ann. Phys. (NY)",
    volume = "470",
    pages = "169790",
    year = "2024"
}

@article{Colladay:1996iz,
    author = "Colladay, Don and Kostelecky, V. Alan",
    title = "{CPT violation and the standard model}",
    eprint = "hep-ph/9703464",
    archivePrefix = "arXiv",
    reportNumber = "IUHET-354",
    doi = "10.1103/PhysRevD.55.6760",
    journal = "Phys. Rev. D",
    volume = "55",
    pages = "6760--6774",
    year = "1997"
}

@article{Kostelecky:2009zp,
    author = "Kostelecky, V. Alan and Mewes, Matthew",
    title = "{Electrodynamics with Lorentz-violating operators of arbitrary dimension}",
    eprint = "0905.0031",
    archivePrefix = "arXiv",
    primaryClass = "hep-ph",
    reportNumber = "IUHET-527",
    doi = "10.1103/PhysRevD.80.015020",
    journal = "Phys. Rev. D",
    volume = "80",
    pages = "015020",
    year = "2009"
}

@article{Casana:2012yj,
    author = "Casana, R. and Ferreira, Jr, M. M. and Silva, E. O. and Passos, E. and dos Santos, F. E. P.",
    title = "{New $CPT$-even and Lorentz-violating nonminimal coupling in the Dirac equation}",
    eprint = "1212.6361",
    archivePrefix = "arXiv",
    primaryClass = "hep-th",
    doi = "10.1103/PhysRevD.87.047701",
    journal = "Phys. Rev. D",
    volume = "87",
    number = "4",
    pages = "047701",
    year = "2013"
}

@article{Magueijo:2001cr,
    author = "Magueijo, Joao and Smolin, Lee",
    title = "{Lorentz invariance with an invariant energy scale}",
    eprint = "hep-th/0112090",
    archivePrefix = "arXiv",
    doi = "10.1103/PhysRevLett.88.190403",
    journal = "Phys. Rev. Lett.",
    volume = "88",
    pages = "190403",
    year = "2002"
}

@article{Breit:1929zz,
    author = "Breit, G.",
    title = "{The Effect of Retardation on the Interaction of Two Electrons}",
    doi = "10.1103/PhysRev.34.553",
    journal = "Phys. Rev.",
    volume = "34",
    pages = "553--573",
    year = "1929"
}

@article{Barut1985DerivationON,
  title={Derivation of Nonperturbative Relativistic Two-Body Equations from the Action Principle in Quantumelectrodynamics},
  author={A. O. Barut and Soliman R. Komy},
  journal={Fortschr. Phys.},
  year={1985},
  volume={33},
  pages={309-318},
  url={https://api.semanticscholar.org/CorpusID:121880433}
}

@article{Salpeter:1951sz,
    author = "Salpeter, E. E. and Bethe, H. A.",
    title = "{A Relativistic equation for bound state problems}",
    doi = "10.1103/PhysRev.84.1232",
    journal = "Phys. Rev.",
    volume = "84",
    pages = "1232--1242",
    year = "1951"
}

@article{Barut1985RadialEF,
  title={Radial Equations for the Relativistic Two-Fermion Problem with the Most General Electric and Magnetic Potentials},
  author={A. O. Barut and Nuri {\"U}nal},
  journal={Fortschr. Phys.},
  year={1985},
  volume={33},
  pages={319-332},
  url={https://api.semanticscholar.org/CorpusID:121270793}
}

@article{Guvendi:2022uvz,
    author = "Guvendi, Abdullah and Zare, Soroush and Hassanabadi, Hassan",
    title = "{Exact solution for a fermion\textendash{}antifermion system with Cornell type nonminimal coupling in the topological defect-generated spacetime}",
    doi = "10.1016/j.dark.2022.101133",
    journal = "Phys. Dark Univ.",
    volume = "38",
    pages = "101133",
    year = "2022"
}

@article{Guvendi:2024nmm,
    author = "Guvendi, Abdullah",
    title = "{Evolution of an interacting fermion\textendash{}antifermion pair in the near-horizon of the BTZ black hole}",
    doi = "10.1140/epjc/s10052-024-12542-x",
    journal = "Eur. Phys. J. C",
    volume = "84",
    number = "2",
    pages = "185",
    year = "2024"
}

@article{Guvendi:2023aor,
    author = "Guvendi, Abdullah and Hassanabadi, Hassan",
    title = "{Fermion-antifermion pair in magnetized optical wormhole background}",
    doi = "10.1016/j.physletb.2023.138045",
    journal = "Phys. Lett. B",
    volume = "843",
    pages = "138045",
    year = "2023"
}

@article{Guvendi:2020kei,
    author = "Guvendi, Abdullah",
    title = "{Relativistic Landau levels for a fermion-antifermion pair interacting through Dirac oscillator interaction}",
    eprint = "2012.04493",
    archivePrefix = "arXiv",
    primaryClass = "physics.gen-ph",
    doi = "10.1140/epjc/s10052-021-08913-3",
    journal = "Eur. Phys. J. C",
    volume = "81",
    number = "2",
    pages = "100",
    year = "2021"
}

@article{Guvendi:2023rex,
    author = "Guvendi, Abdullah and Mustafa, Omar",
    title = "{Fermion-antifermion pairs in a magnetized space-time with non-zero cosmological constant}",
    eprint = "2401.02441",
    archivePrefix = "arXiv",
    primaryClass = "physics.gen-ph",
    doi = "10.1016/j.nuclphysb.2024.116571",
    journal = "Nucl. Phys. B",
    volume = "1004",
    pages = "116571",
    year = "2024"
}

@article{Dogan:2023lil,
    author = "Dogan, Semra Gurtas",
    title = "{Dirac pair in magnetized elliptic wormhole}",
    doi = "10.1016/j.aop.2023.169344",
    journal = "Ann. Phys. (NY)",
    volume = "454",
    pages = "169344",
    year = "2023"
}

@article{Guvendi:2024diq,
    author = "Guvendi, Abdullah and Mustafa, Omar",
    title = "{An innovative model for coupled fermion-antifermion pairs}",
    eprint = "2405.16290",
    archivePrefix = "arXiv",
    primaryClass = "hep-ph",
    doi = "10.1140/epjc/s10052-024-13192-9",
    journal = "Eur. Phys. J. C",
    volume = "84",
    number = "8",
    pages = "866",
    year = "2024"
}

\end{document}